\documentclass[pra,twocolumn,showpacs,preprintnumbers,amsmath,amssymb]{revtex4-1}

\usepackage{graphicx}
\usepackage{dcolumn}
\usepackage{bm}
\usepackage{psfrag}
\usepackage{epsfig}
\usepackage{amsmath}
\usepackage{amssymb}
\usepackage{color}
\usepackage{bbm}
\usepackage[FIGTOPCAP,raggedright,nooneline]{subfigure}
\usepackage{verbatim}
 \usepackage[applemac]{inputenc}

\usepackage{tikz}
\newcommand*\circled[1]{\tikz[baseline=(char.base)]{
            \node[shape=circle,draw,inner sep=1pt] (char) {#1};}}

\newcommand{\ignore}[1]{}

\newcommand{\sgn}{\text{sgn}}

\begin{document}

\title{Strong coupling between moving atoms and slow-light Cherenkov photons}

\author{Giuseppe Calaj\'o and Peter Rabl}
\affiliation{Vienna Center for Quantum Science and Technology,
Atominstitut, TU Wien, 1040 Vienna, Austria}

\date{\today}

\begin{abstract}
We describe the coupling of moving atoms to a one dimensional photonic waveguide in the regime where the atomic velocities are comparable to the effective speed of light. Such conditions could be achieved, for example, in photonic crystals or coupled resonator arrays, where the maximal photonic group velocity is significantly reduced compared to free space. In this case the interplay between a velocity-induced directionality and the emergence of new divergencies in the photonic density of states gives rise to a range of novel phenomena and non-perturbative effects in the emission of photons and the resulting photon-mediated interactions between moving atoms.   
We show that apart from potential implementations with optical waveguides, Rydberg atoms flying above a coupled array of superconducting microwave resonators provide a versatile platform for exploring this new regime of atom-light interactions under experimentally accessible conditions. 
\end{abstract}

\maketitle

%
%

\section{Introduction}

The radiation of charged particles moving close to the speed of light is a long-studied problem of classical electrodynamics~\cite{Jack} 
with practical relevance for accelerator physics or the development of intense X-ray sources~\cite{FEL}.
In contrast, the coupling of neutral atoms and molecules to classical or quantized fields is usually investigated under opposite conditions, where particle velocities are low and mainly affect the emission and absorption of photons via a small Doppler shift of the transition frequency. This effect, however, is the basis for laser cooling~\cite{StenholmRMP1986} and successive atom trapping techniques~\cite{Metcalf}, which enabled the study of strong light-matter interactions even at the single photon level~\cite{BrinbaumNature2005,TieckeNature2014,Volz2014,HackerNature2016}. Based on the same techniques, it has recently become possible to interface cold atoms with one dimensional (1D) waveguides~\cite{ReitzPRL2013,YallaPRL2014} and photonic crystal structures~\cite{ThompsonScience2013,GobanNatComm2014}, inside which the propagation velocity of photons can be significantly reduced~\cite{Krauss}. This opens up new  possibilities for exploring an intriguing regime of atom-light interactions, where individual atoms and photons move at comparable velocities, while still interacting strongly with  each other.

   
The coupling of atoms or other emitters~\cite{review-lodahl} to 1D  waveguides, photonic crystals or coupled-cavity arrays has already attracted considerable attention. In this context, many interesting phenomena like single and multi-photon scattering~\cite{FanPRL2007,ZhouPRL2008,Sun2,LongoPRL2010}, atom-photon bound states~\cite{Byk,John1990,John1994,Kofman,lambro,LongoPRL2010,Palma,calajo,Tao,Houck}, self-organization~\cite{ChangPRL2013,GriesserPRL2013} and quantum many-body effects~\cite{Hartmann2006, GreentreeNatPhys2006, AngelakisPRA2007}, or unidirectional emission~\cite{PetersenScience2014,MitschNatComm2014,SollnerNatNano2015,Stannigel,ChiralRev}
 have been explored, usually under the assumption of stationary or slowly moving emitters. In this work we go beyond this conventional scenario and study atom-photon interactions in slow-light waveguides under conditions, where the atomic velocities approach, match or even exceed the effective speed of light. We show that the existence of an upper bound for the photonic group velocity  
introduces different types of directionalities and novel non-perturbative effects in photon emission and photon-transfer processes, which are not present for static atoms. In particular, we describe the appearance of  a velocity-induced divergence in the photonic density of states, which is associated with a strong resonant coupling between the moving atoms and individual co-propagating Cherenkov photons. 

\begin{figure}
\includegraphics[width=0.48\textwidth]{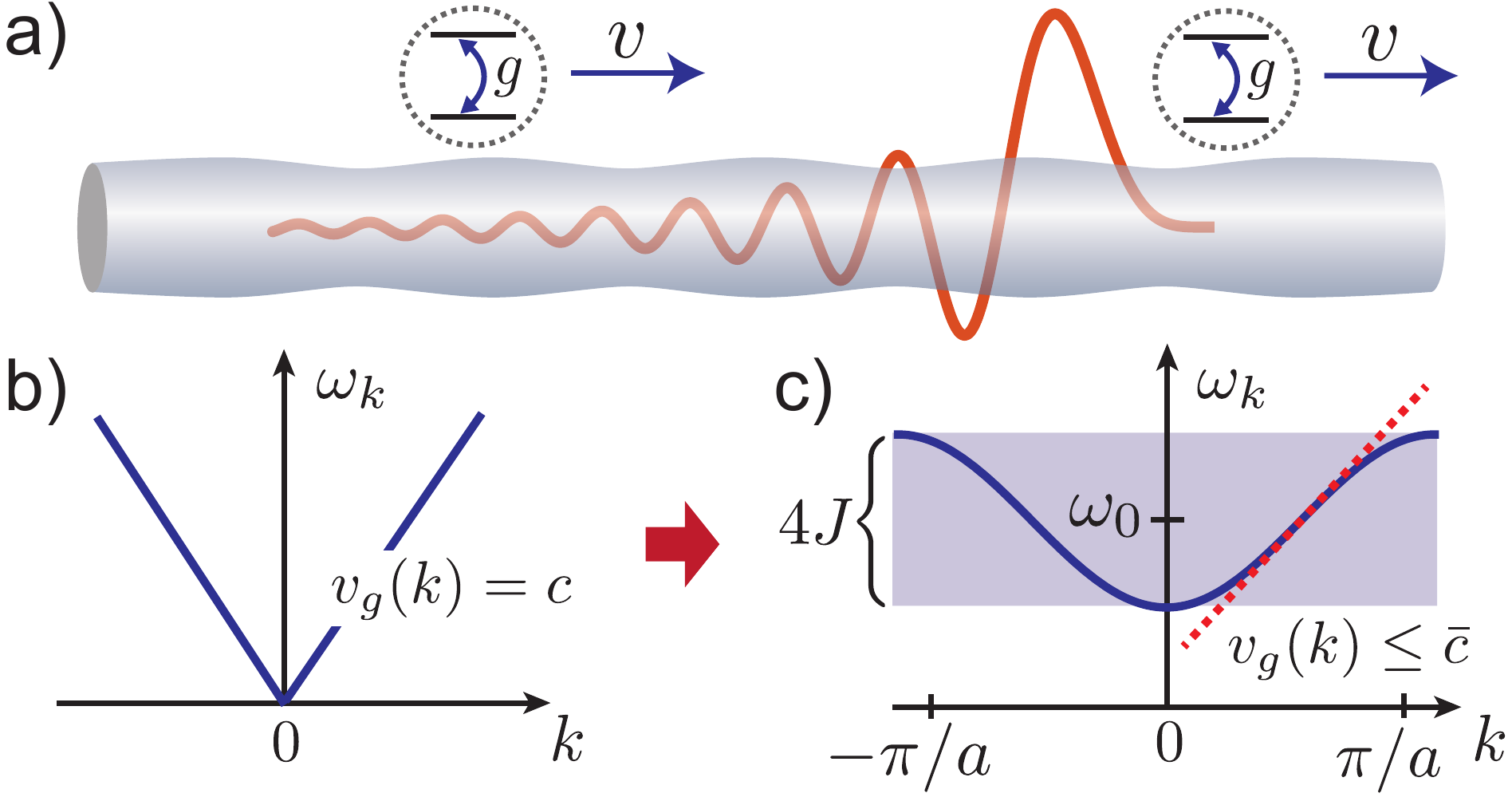}%
\caption{Setup. a) Two-level atoms moving at a velocity $v$ are strongly coupled to the evanescent field of a 1D waveguide. By introducing a  sufficiently strong spatial modulation with periodicity $a$, photons are slowed down by Bragg-reflection and the usual linear dispersion relation shown in b) separates into individual bands of finite width $4J$, as indicated in c). Within one band the maximal group velocity  is bounded by an effective speed of light, $\bar c=2Ja$, which can be comparable to typical atomic velocities.
}
\label{fig:Setup}
\end{figure}

The new regime of light-matter interactions addressed in this work could potentially be realized with flying atoms coupled to various optimized photonic waveguide and photonic crystal structures. To overcome some of the experimental challenges of implementing such systems in the optical domain, we further describe a new realization of a slow-light waveguide QED system consisting of Rydberg atoms flying across a coupled array of planar microwave resonators~\cite{petro,Hogan,HermannPRA2014,BeckAPL2016,UnderwoodPRA2012}. This setting provides a versatile experimental platform for exploring all of the here described phenomena and to study, for example, classical effects like Cherenkov radiation~\cite{Cherenkov,Jelly,Luo, Luo2} in a quantum optical context, where each individual photon is interacting strongly with the moving atoms. In this setup the velocity-enhanced excitation transfer efficiency between atomic and photonic states could also play a role for developing and optimizing new coherent interfaces between quantum optical and solid-state systems~\cite{XiangRMP2013,KurizkiPNAS2015}, using fast moving atoms.

\section{Model}

We consider a setup as shown in Fig.~\ref{fig:Setup} a), where $N$ two-level atoms are coupled to the field of a 1D photonic waveguide.  The atoms have a mass $m$ and the internal ground state $|g\rangle$ and excited state $|e\rangle$ are separated in frequency by $\omega_a$. Restricted to the atomic motion along the waveguide direction, $\vec e_z$, the generic Hamiltonian for this system reads
\begin{equation}\label{H_general}
\begin{split}
H=&\sum_{i=1}^{N}\left(\frac{p_{i}^2}{2m}+\hbar \omega_a|e\rangle_i\langle e|\right)+\int dk  \, \hbar \omega_k \, a_k^{\dagger}a_k\\
+&\hbar \sum_{i=1}^{N} \int \frac{dk}{\sqrt{2\pi}} \left(g_k e^{ikz_i}a_k\sigma^i_+ + g_k^* e^{-ikz_i}a^\dag_k\sigma^i_- \right).
\end{split}
\end{equation}
Here the $z_i$ and $p_i$ are the atomic position and momentum operators, $\sigma^i_{-}=(\sigma^i_{+})^{\dagger}=|g\rangle_i\langle e|$, and the $a_k$, where $[a_k,a_{k'}^{\dagger}]=\delta(k-k')$, are the bosonic annihilation operators for photons with wavevector $k$. In conventional waveguides and within the relevant frequency range, we usually obtain $g_k\simeq const.$ and $\omega_k\simeq (c/n)|k|$, where $c$ is the speed of light in vacuum  and $n\sim\mathcal{O}(1)$ is the refractive index of the material. Therefore, in this case the photonic group velocity, $v_g(k)=\partial \omega_k /\partial k\sim 10^8$ m/s, is much larger than typical velocities $v\lesssim 10^4$ m/s of neutral atoms.  

\subsection{Slow-light waveguides}
To achieve a condition $v\sim v_g$, we consider in this work a waveguide with a spatially periodic modulation (for example, of its width, refractive index, etc.) with lattice constant $a$ and total length $L\gg a$. Provided that this modulation is sufficiently strong, the propagation of photons is slowed down by Bragg-reflection~\cite{Krauss} and the spectrum separates into individual bands, each with an approximate dispersion relation  of the form
\begin{equation}\label{eq:Dispersion}
\omega_{k} \simeq  \omega_{0} - 2J \cos(ka),
\end{equation}
where $\omega_{0}$ is the central frequency and $4J$ the total width of the band [cf. Fig.~\ref{fig:Setup} c)].  By assuming that all the relevant timescales of the system dynamics are slow compared to the inverse frequency separation between individual bands, $\sim a/c$, we can restrict our analysis to a single band and we  obtain 
\begin{equation}\label{H discrete}
\begin{split}
&H=\sum_{i=1}^{N}\left(\frac{p_{i}^2}{2m}+\hbar \omega_a|e\rangle_i\langle e|\right) +\sum_{k} \hbar\omega_{k}  \, a_{k}^{\dagger}a_{k}\\
&+\sum_{i=1}^{N}\sum_k  \hbar g  \left[  \psi_{k}(z_i)a_k\sigma^i_+ +  \psi^*_{k}(z_i) a_k^{\dagger}\sigma^i_- \right].
\end{split}
\end{equation}
Here the  $a_k$ and $a_k^\dag$, obeying $[a_k,a_{k'}^{\dagger}]=\delta_{k,k'}$, are now bosonic annihilation and creation operators for photons with  Bloch-wavefunction
\begin{equation}
\psi_{k}(z)= \sqrt{\frac{a}{L}} u_{k}(z)  e^{i k z},
\end{equation}
where $k\in (-\pi/a,\pi/a]$ and the $u_{k}(z)=u_{k}(z+a)$ are periodic functions. 

From Eq.~\eqref{eq:Dispersion} we see that within the single band of interest, the  periodic modulation  introduces an upper bound for the maximal propagation velocity of the photons given by the effective speed of light 
\begin{equation}
\bar c = {\rm max} \{v_g(k)\} = 2Ja.
\end{equation}
 It can be tuned close to zero by taking the limiting case of a weakly tunnel-coupled array of localized resonator modes, making the regime $v \sim \bar c$ experimentally accessible. Note, however, that the required reduction of the photonic group velocity by many orders of magnitude goes significantly beyond what is possible in conventional photonic crystal structures~\cite{Krauss}. In Sec.~\ref{Implementation} below we will discuss in more detail, how this extreme slow-light condition can be realized in practice.

\subsection{Continuum model}
It is important to keep in mind that by introducing a periodic modulation for slowing down the photons, we have changed from a continuum to a Bloch-wave description, which also results in a  periodic variation of the coupling, $g\rightarrow g(z)\sim u(z)$. For atoms moving along classical trajectories, i.e.,  $z_i(t)= z_{i}+v_i t$, this translates into a time-periodic modulation of the coupling with frequency $\Omega_i=2\pi |v_i|/a$, which makes the dynamics of slow-light systems in general quite involved. However, it turns out that for a large range of parameters, in particular for the relevant case of high velocities, $v\sim \bar c$, and atomic frequencies $\omega_{a}$ inside the propagation band, we can derive an effective model [see App.~\ref{AppA}]
\begin{equation}\label{H_tot}
\begin{split}
&H(t)\simeq \sum_{i=1}^{N} \hbar \omega_a |e\rangle_i\langle e| +\sum_{k} \hbar\omega_k  \, a_k^{\dagger}a_k\\
&+\sum_{i=1}^{N} \sqrt{\frac{a}{L}} \sum_{k}\hbar\bar g \left[ e^{ikz_i(t)}a_k\sigma^i_+ + e^{-ikz_i(t)}a_k^\dag \sigma^i_- \right],
\end{split}
\end{equation}
 where only the coupling averaged over one unit cell,
 \begin{equation}\label{gbar}
\bar g= g\times \frac{1}{a} \int_{0}^{a} dz\,  u(z),
\end{equation}
 appears. This model now mimics very closely the coupling of fast moving atoms to arbitrarily slow photons in a 1D continuum and will be used as a starting point for the following analysis. A more detailed discussion of the range of  validity of Eq.~\eqref{H_tot}  is given in Sec.~\ref{Validity} below.

\section{Atoms and photons interacting at the speed of light}
\begin{figure}
 \includegraphics[width=0.52\textwidth]{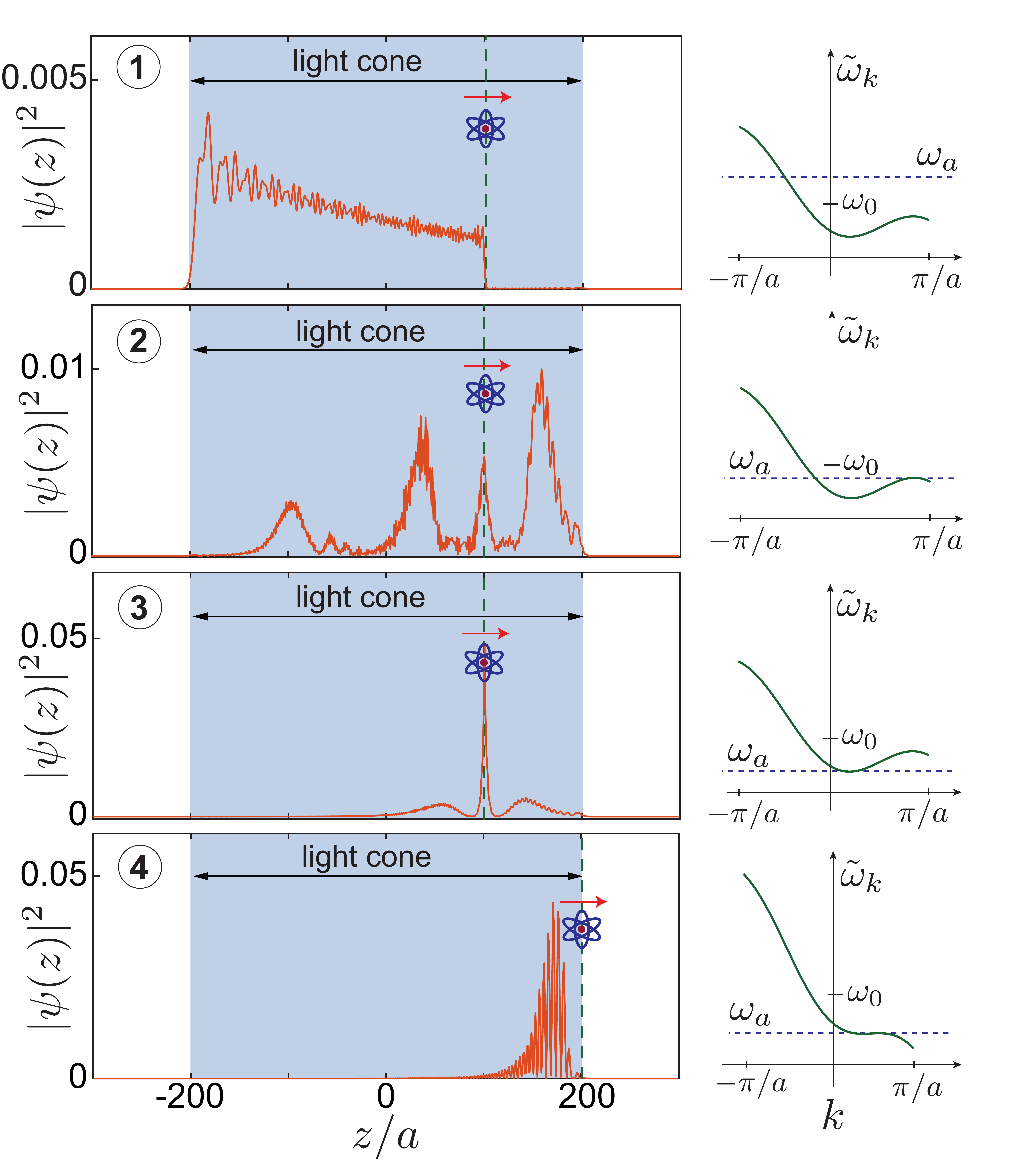}
 \caption{ Snapshots of the emitted photon wavepacket evaluated at a time $t=100/J$ and for the parameters \protect\circled{1} $v=\bar c/2$, $\delta/(2J)=1$, \protect\circled{2} $v=\bar c/2$,  $\delta/(2J)=-0.5$, \protect\circled{3} $v=\bar c/2$, $\delta/(2J)=-1.1$ and \protect\circled{4} $v=\bar c$, $\delta/(2J)=-\pi/2$. The blue shaded area indicates the light cone, $|z| \leq \bar c t$.  For each set of parameters the plots in the right column show the position of the atomic transition frequency within the tilted photonic propagation band in the co-moving frame.}
\label{fig:EmissionA}
\end{figure}

Based on the effective continuum model given in Eq.~\eqref{H_tot}, we now investigate in this section, how basic atom-photon processes are modified when atoms and photons move at comparable velocities.  
We first consider the emission of a photon from a single atom moving at a constant velocity $v>0$. By assuming that at time $t=0$ the atom is  in state $|e\rangle$ and the waveguide in the vacuum state $|{\rm vac}\rangle$, the state of the whole system can be written as $|\psi(t)\rangle= \big[c_e(t) \sigma_++ \sum_k \, \psi(k,t) a^\dag_k\big]|g\rangle|{\rm vac}\rangle$, where $p_e(t)=|c_{e}(t)|^2$ is the excited state population and $\psi(k,t)$ is the wavefunction of the emitted photon in $k$-space. Figure~\ref{fig:EmissionA} shows snapshots of the resulting wavefunction in the discretized position space, $\psi(z,t)$, for different ratios $v/\bar c$ and different atom-waveguide detunings $\delta=\omega_a-\omega_0$.

\subsection{Velocity-induced directional emission of light}

To obtain a better intuition for these emission patterns, it is convenient to change into a frame, which is co-moving with the atom. This is achieved by the unitary transformation $\tilde H= UHU^\dag+i\hbar\dot U U^\dag$, where $U(t)=e^{i\sum_kvk a_k^{\dagger}a_kt}$. In this new representation the time-dependence in the exponentials in Eq.~\eqref{H_tot} is eliminated, $z(t)\rightarrow z$,  but we obtain instead a tilted dispersion relation $\omega_k\rightarrow \tilde \omega_k =\omega_k-kv$. For frequencies $\omega_a$ inside this tilted band and for sufficiently small $\bar g$, the coupling to the continuum of modes in the waveguide will result in an approximately exponential decay of $p_e(t)$ with total rate $\Gamma=\Gamma_L+\Gamma_R$, where 
\begin{equation}\label{eq:GammaLR}
\Gamma_{L,R} = \frac{\bar g^2 a }{| \tilde v_g (k_{L,R})|}.
\end{equation}
Here $k_L<0$ and $k_R>0$ are left- and right-propagating wavevectors defined by the resonance condition $\tilde \omega_k = \omega_a$, and $\tilde v_g(k)=v_g(k)-v$ is the  group velocity in the co-moving frame. In the usual case, where $v_g(k)\simeq \pm \bar c$ and $v\ll \bar c$, Eq.~\eqref{eq:GammaLR} implies that photons are emitted with Doppler-shifted wavevectors $k_{L,R}=\mp \omega_a/(\bar c \pm v)$, but with the same group velocity in the laboratory frame and approximately the same rates $\Gamma_L\simeq \Gamma_R$.  

\begin{figure}
 \includegraphics[width=0.45\textwidth]{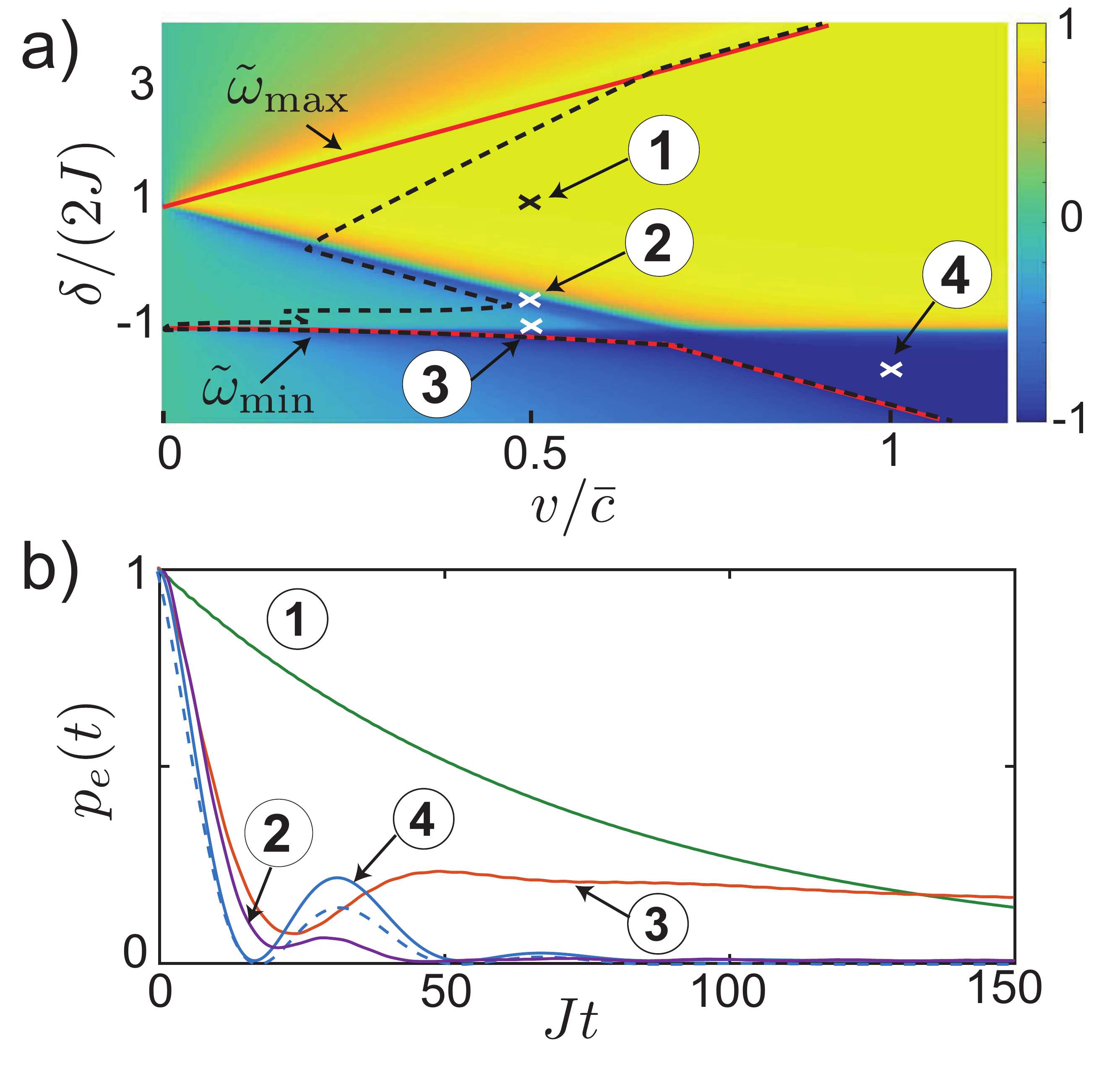}
 \caption{a) Plot of the directionality parameter $D$ defined in Eq.~\eqref{eq:Directionality}  as a function of $\delta$ and $v$. The crosses mark the parameter values assumed in the corresponding plots in Fig.~\ref{fig:EmissionA}. The solid red lines indicate the minimal ($\tilde \omega_{\rm min}$) and maximal ($\tilde \omega_{\rm max}$) frequency of the tilted band. For each detuning, the dashed line shows the minimal velocity $v_{\rm min}(\delta)$, above which the effective model~\eqref{H_tot} provides an accurate description of the dynamics (see  Sec.~\ref{Validity} for more details). b) Decay of the excited state population $p_e(t)$ for different sets of parameters. The dashed line indicates the approximate analytic result given in Eq.~\eqref{pop_approx} for $v=\bar c$ and $\delta=-\pi J$. In all plots a coupling of $\bar g/(2J)=0.1$ is assumed. }
\label{fig:EmissionB}
\end{figure}

In the regime $v\sim \bar c$, this picture changes significantly. As illustrated in example \circled{1} in Fig.~\ref{fig:EmissionA}, the tilting of the finite propagation band opens a frequency window, within which only backward propagating modes exist. This gives rise to a purely unidirectional emission opposite to the atomic motion.  It is interesting to see that this fully directional emission already occurs at velocities $v<\bar c$, where photonic wavepackets traveling faster than the atom would still be allowed. This is indeed the case for the parameters chosen in example \circled{2}, where a different mechanism for directionality sets in. For this value of the atomic frequency the emission is enhanced in forward direction due to a divergence in the photonic density of states, i.e. $\tilde v_g(k_R)\simeq 0$, associated with the right edge of the tilted propagation band. Although emission into backward-propagating modes is still allowed, this band-edge effect leads to a strong asymmetry, $\Gamma_R\gg\Gamma_L$, as well as a substantially increased total emission rate. Overall, the combination of both of these mechanisms, related to either the absence or divergence of the photonic density of states,  can give rise to strong variations in the direction, the rate and the group velocity of the emitted photons. This is summarized in Fig.~\ref{fig:EmissionB} a), where we plot the directionality parameter~\cite{ChiralRev,Directionality}
\begin{equation}\label{eq:Directionality}
 D=\frac{\Gamma_L-\Gamma_R}{\Gamma_L+\Gamma_R},
\end{equation} 
for a range of atomic velocities and detunings. Note that for this plot we have assumed a finite decay rate $\gamma_{p}/J=0.01$ for the photons, which allows us to generalize Eq.~\eqref{eq:GammaLR} to frequencies outside the band  and to avoid unphysical divergencies~\cite{calajo}.

\subsection{Atom-photon bound states at finite velocities}

The exponential decay with rates $\Gamma_{L,R}$ given in Eq.~\eqref{eq:GammaLR} assumes a perturbative treatment of the photon emission process, which is strictly valid only in the weak-coupling limit $\bar g \ll |\tilde v(k_{L,R})|/a$. For static atoms it is well known that even for $\bar g \ll J$, such a description breaks down near the edges of the propagation band, where a vanishing group velocity results in a highly non-Markovian, oscillatory decay~\cite{John1994,Kofman} and a non-decaying photonic fraction. This behavior can be understood from the emergence of atom-photon bound states~\cite{John1990}, which are energetically separated from the propagation band. Examples \circled{2} and \circled{3} in Fig.~\ref{fig:EmissionA} and in Fig.~\ref{fig:EmissionB} b)  show that such effects prevail even at very high velocities. In both examples, a large fraction of the photon is coherently reabsorbed and remains bound to the moving atom for a long time. 

We emphasize that in contrast to the static case, the photons in these examples are only quasi-bound to the moving atoms and will eventually decay at very long times.  In example \circled{2}, this decay can already be understood from the tilted dispersion relation shown in Fig. \ref{fig:EmissionA}. Here, apart from the strong coupling to right-moving photons near the band edge, $k\approx \pi/2>0$, the atom can decay as well via the coupling to the continuum of left-propagating modes. For $\omega_a$ near the minimal and maximal frequencies of the tilted band, $\tilde \omega_{\rm min}$ and  $\tilde \omega_{\rm max}$, such a decay channel is absent. Indeed, similar to the case of static atoms~\cite{John1990,John1994,Kofman,lambro,Palma,calajo},  the eigenvalue equation for the single-excitation subspace of Hamiltonian~\eqref{H_tot} in the co-moving frame,
\begin{equation}
\omega_{\pm}-\delta=\frac{\bar g^2a}{2\pi}\int_{-\pi/a}^{\pi/a}\frac{d k}{ \omega_{\pm}-\tilde\omega_k} ,
\end{equation}
always predicts two non-decaying bound states with eigen-frequencies $\omega_{\pm}$ outside the propagation band. However, such an analysis would also predict the existence of atom-photon bound states at velocities $v>\bar c$, which are clearly unphysical. As we discuss in more detail below, these paradox predictions are a consequence of the continuum approximation used in Eq.~\eqref{H_tot}. They disappear when the full spatial dependence of $g(z)$ is taken into account and bound states can decay via additional channels at the sideband frequencies $\omega_{a}\pm \Omega$~\cite{Luo2}. We emphasize though that all the results shown in Fig.~\ref{fig:EmissionA}, including in particular the quasi-bound state in example \circled{2}, are not artifacts of the continuum approximation and are accurately reproduced by the full model on the timescales of interest.

\subsection{Photon emission at the speed of light}
In addition to the above mentioned band-edge effects, which are already present at low atomic velocities~\cite{Luo2}, a new and very unique situation occurs when the atomic velocity matches the speed of light, $v=\bar c$ [see example \circled{4} in Fig.~\ref{fig:EmissionA}) and  Fig.~\ref{fig:EmissionB} b)]. In this case one obtains a cubic dispersion relation 
\begin{equation}\label{cubic}
\tilde \omega_k \simeq -\pi J -\frac{Ja^3}{3} (k-k_c)^3,
\end{equation}
 for wavevectors around $k_c=\pi/(2a)$.  
This results in a higher order divergence $\sim (\omega-\pi J)^{-2/3}$ in the effective photonic density of states. However, in contrast to usual band-edge effects, this velocity-induced singularity lies in the middle of the propagation band and for moderate $\bar g$ there is no bound state associated with it.  This divergence results in  strong interactions between the moving atom and the co-propagating Cherenkov photon with group velocity $v_{g}(k_c)= v$.

In App.~\ref{AppB} we present a detailed analysis of the single photon eigenstates for a simplified system consisting of a single atom coupled to a photonic waveguide with a purely cubic dispersion relation. From these analytic results we obtain  the following approximate shape for the atomic decay [see Fig~\ref{fig:EmissionB} b)],
\begin{equation}\label{pop_approx}
p_e(t) \simeq \cos^2(\Omega_{c} t ) e^{-\Gamma_{c} t}, 
\end{equation}
where coherent atom-photon oscillations and the overall decay occur with very similar characteristic rates \begin{equation}\label{coefficients}
\Omega_c=\frac{\sqrt{5+\sqrt{5}}}{2\sqrt{2}}\left(\frac{\bar g^6}{9J}\right)^{\frac{1}{5}},\qquad \Gamma_c=\frac{\sqrt{5}-1}{2}\left(\frac{\bar g^6}{9J}\right)^{\frac{1}{5}}.
\end{equation} 
The oscillatory dependence of $p_e(t)$ and the unusual scaling, $\Omega_c,\Gamma_c\sim \bar g^{6/5}$, demonstrate that atom-light interactions in this critical parameter regime are highly non-perturbative and show a clear strong coupling behavior already at the level of individual photons. 
For a given $\bar g \ll J$, $\Gamma_c$ is also the fastest rate at which the atom can be completely deexcited by emitting a highly localized photonic wave packet, which closely follows the atom in forward direction. 

At even higher velocities, $v>\bar c$, the density of states, and therefore also the emission rate, decreases again. As shown in Fig.~\ref{fig:EmissionB} a), the directionality of the emitted photon in the laboratory frame is then only determined by the condition $\delta \lessgtr -2J$.  For the specific detuning $\delta=-2J$, a very extended and essentially stationary photonic wavepacket  is produced.

\begin{figure}
\includegraphics[width=0.48\textwidth]{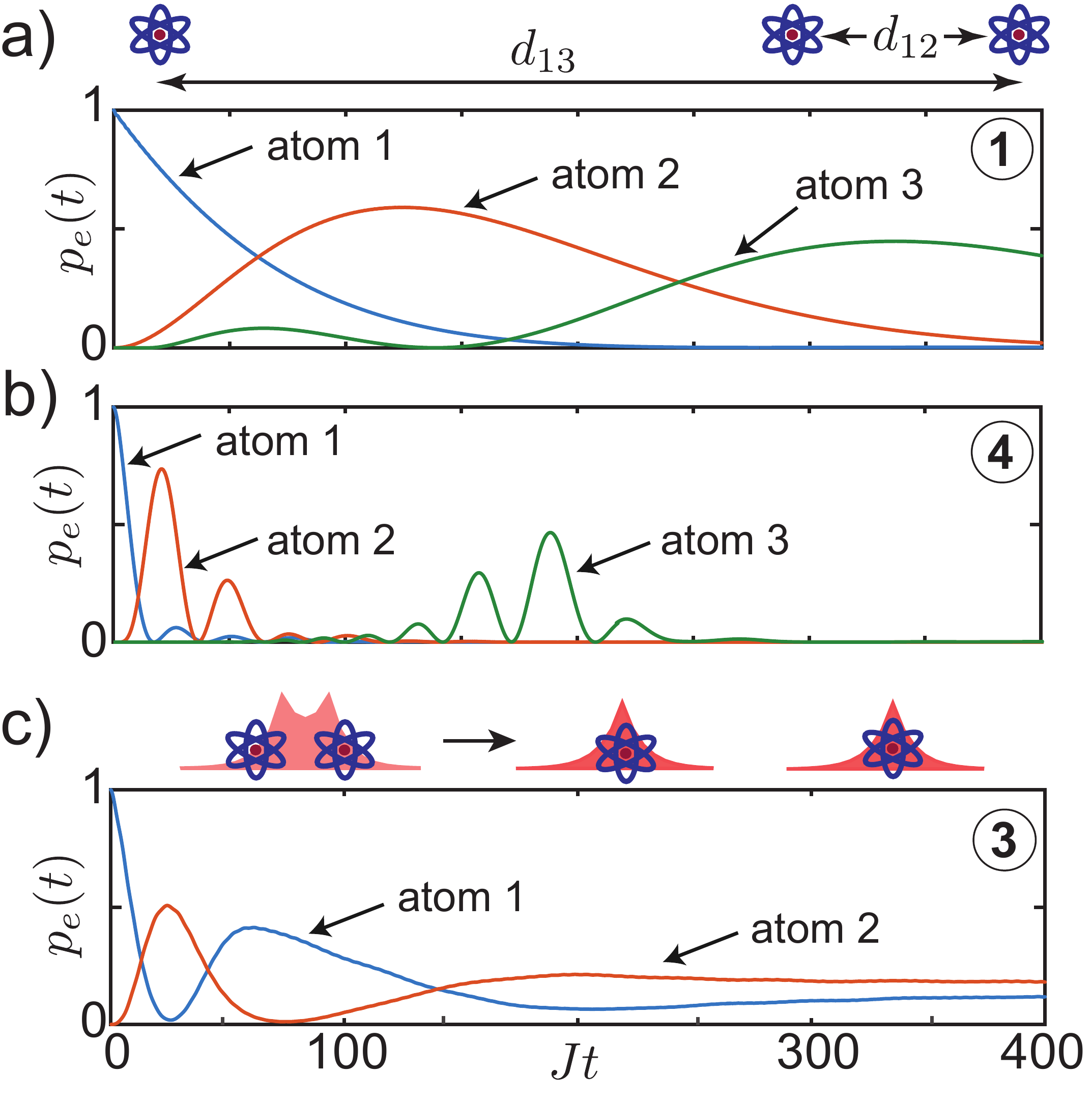}%
\caption{Excitation transfer between moving atoms. In the first two plots three atoms moving with the same velocity $v>0$ and relative separations $d_{12}/a=2$ and $d_{13}/a=45$ are considered.  The parameters in  a) are $v=\bar c/2$ and $\delta/(2J)=1$ and in b) $v=\bar c$ and $\delta=-\pi J$. For the plot in c) two atoms with $z_1(0)=z_2(0)$, slightly different velocities $v_1=0.5\bar c$ and $v_2=0.53\bar c$  and a detuning of $\delta/(2J)=-1.1$ have been assumed. In all plots $\bar g/(2J)=0.1$.}
\label{fig:Transfer}
\end{figure}

\subsection{Excitation transfer processes}
Compared to conventional Cherenkov radiation studied in higher dimensional structures~\cite{Jelly,Luo, Luo2}, a key feature of the current setting is that all emitted photons are confined to one dimension and can thus be efficiently reabsorbed by other atoms. For conventional waveguides, this emission and reabsorption of photons gives rise to almost instantaneous, long-range and bidirectional dipole-dipole interactions, which depend only on the relative atomic positions, $|z_i-z_j|$~\cite{ChangNJP2012,GonzalesTudelaPRL2013}.  In the extreme slow-light regime various qualitatively different excitation transfer processes arise, depending strongly on the velocities and detunings of the involved atoms. To illustrate this point we first consider in Fig.~\ref{fig:Transfer} a) and b) the case of $N=3$ atoms moving at the same speed $v>0$, with the first atom to the right being initially prepared in the excited state. For this plot we have solved the evolution of the generalized wavefunction $|\psi(t)\rangle= \big[\sum_{i=1}^N c^i_e(t) \sigma^i_++ \sum_k \, \psi(k,t) a^\dag_k\big]|g\rangle|{\rm vac}\rangle$ under the action of Hamiltonian~\eqref{H_tot}. 

With all the atoms at rest and under the usual Markov approximation, where the propagation time of the photons can be neglected, the maximal population transfer from atom 1 to atoms 2 and 3 is limited to $p_e^{(2,3)}(t)<0.25$. This result can be understood from a decomposition of the initial atomic excitation into super- and sub-radiant states~\cite{Longhi,Pascazio,Moreno} and depends in detail on the exact atomic positions. This limit no longer applies at moderate and fast velocities, where the emission becomes highly directional and therefore also results in a much more efficient transfer of excitations from the right to the left. This is shown in  Fig.~\ref{fig:Transfer} a), where the same parameters as in example \circled{1} in Fig.~\ref{fig:EmissionA} for a completely unidirectional photon emission have been assumed.  
Note that for these parameters the retardation time $\tau_{13}=|d_{13}/(v_g(k_L)-v)|$   is still short compared to the timescale of the transfer dynamics $(\sim \Gamma_L^{-1})$ and the whole process can  be well described by a unidirectional Markovian master equation~\cite{Stannigel}  in the co-moving frame. This is no longer the case for the critical condition $v=\bar c$ and $\delta/J=-\pi$ assumed in Fig.~\ref{fig:Transfer} b). Here we observe a very rapid transfer of the initial excitation from atom 1 to atom 2, which is mediated by the highly compressed Cherenkov photon shown in example \circled{4} in Fig.~\ref{fig:EmissionA}.  This photon then slowly falls behind and at a much later time overlaps with the third atom. The resulting oscillations between the photon and atom 3 with roughly the same Rabi frequency $\Omega_c$ as identified in Eq.~\eqref{pop_approx}, are again a clear signature of the highly non-Markovian nature of the transfer process.

Finally, in Fig.~\ref{fig:Transfer} c) we consider another example, where two initially co-propagating atoms exchange excitations via a co-moving bound photon, which remains exponentially localized around the atomic positions $z_i(t)$.  By introducing a small velocity difference, i.e. $v_1\neq v_2$, the atoms slowly separate and gradually decouple. Depending on the detailed choice of parameters, the photon may remain with one of the atoms or be split into two separated atom-photon bound states~\cite{calajo}.

\subsection{Validity of the effective continuum model}\label{Validity}
\begin{figure}
\includegraphics[width=0.48\textwidth]{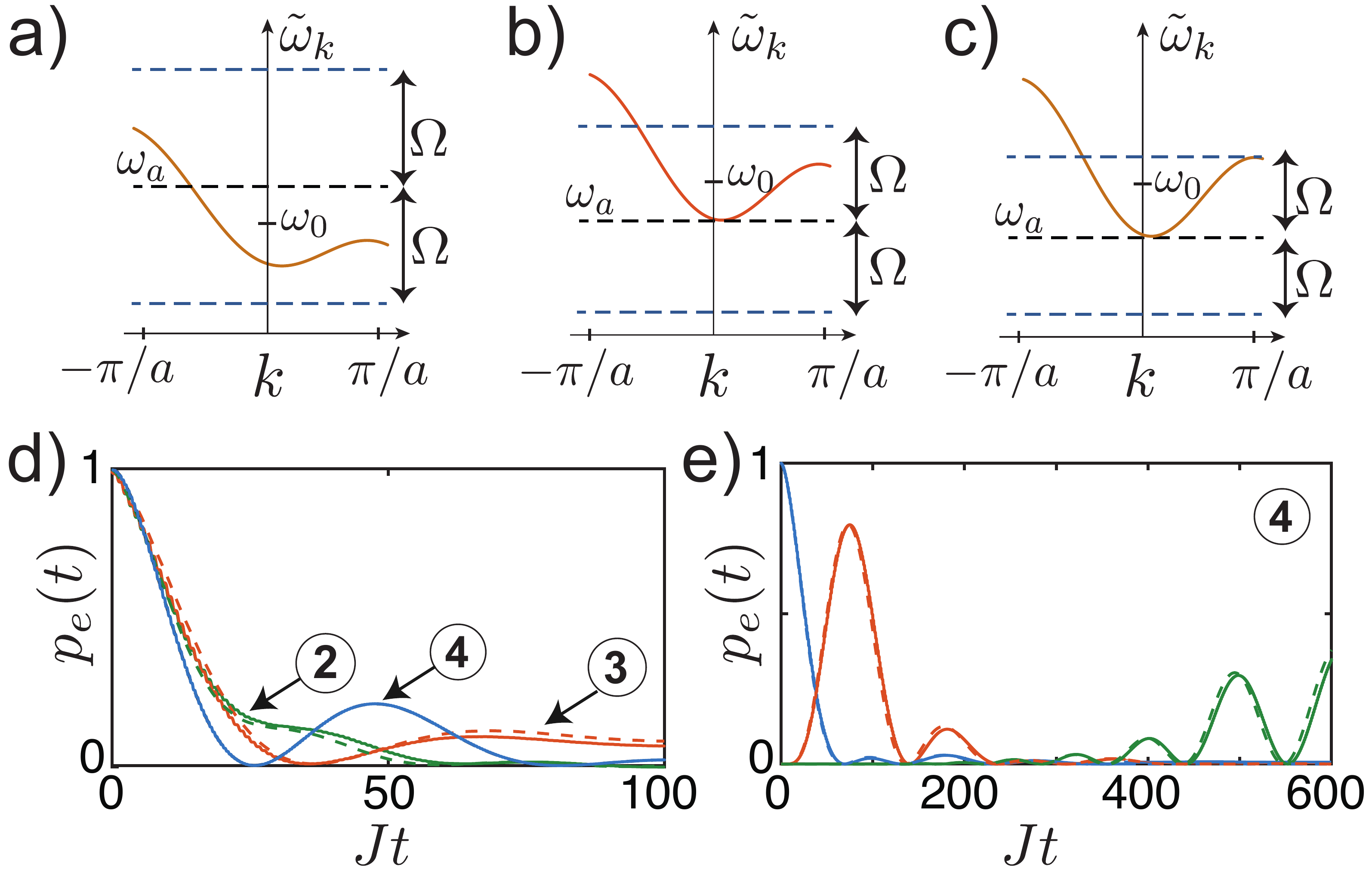}%
\caption{ a)-c) The positions of the modulation frequencies $\omega_a\pm\Omega$ are shown in relation to the tilted photonic band for a) $v/\bar c=0.5$ and $\delta/(2J)=1$,  b) $v/\bar c=0.3$ and $\delta/(2J)=-1.05$ and c) $v/\bar c=0.22$ and $\delta/(2J)=-1$. In d) and e) the evolution of the excited state population as predicted by the full model in Eq.~\eqref{H discrete} (continuous line) is compared with the corresponding evolution under the effective continuum model in Eq.~\eqref{H_tot} (dashed line). 
In all the plots $g/(2J)=0.1$ and $z_0/a=0.1$ is assumed. All the other parameters are the same as in the respective examples in Fig.~\ref{fig:EmissionA} and Fig.~\ref{fig:Transfer}.
 }\label{figsup1}
 \end{figure}

Our discussion so far has been based in the approximate continuum model given in Eq.~\eqref{H_tot}, which ignores the time-periodic modulation of the coupling with multiples of the frequency $\Omega=2\pi |v|/a$. To understand more precisely, under which conditions this description is meaningful, we plot in Fig.~\ref{figsup1} a)-c) the tilted dispersion relation $\tilde \omega_k$ in relation to the atomic frequency $\omega_a$ and the first two modulation sidebands at $\omega_a\pm \Omega$. As described in more detail in App. A, the modulation of the coupling can induce additional decay channels with rates $\Gamma^{(n)}$ proportional to the density of states evaluated at the modulation frequencies $\omega_a-(\omega_k-vk)+n \Omega$, where $n=\pm1,\pm2, ... $~\cite{Luo2}. To ensure that such corrections do not affect the dynamics, all sidebands $\omega_a + n \Omega$ must lie outside the tilted propagation band, as illustrated in Fig.~\ref{figsup1} a). This condition can be recast into the form 
\begin{equation}\label{condition_main}
\Omega=\frac{2\pi |v|}{a} > {\rm max}_k\left|\frac{\delta+2J\cos(ka)}{1-\frac{ak}{2\pi}}\right|,
\end{equation}
and implies that a continuum description for a slow-light waveguide is only possible for high enough velocities, $|v|\gtrsim 0.25 \bar c$, and for atomic frequencies inside the tilted propagation band. This simple picture also explains why there are no bound state, e.g., in the gap above the upper band edge, $\omega_+\gtrsim \tilde \omega_{\rm max}=\omega_0+2J+\Omega/2$. In this case the sideband $\omega_+-\Omega$ is necessarily inside the propagation window and leads to a rapid decay. 

While Eq.~\eqref{condition_main} captures very well the main region of validity of Eq.~\eqref{H_tot}, it is too strict in certain regimes. In particular, this is the case for moderately slow atoms tuned to the lower band edge, $\omega_a\simeq \tilde\omega_{\rm min}$. As indicated in Fig.~\ref{figsup1} b), for these parameters the condition~\eqref{condition_main} can be violated, but the dynamics is still well approximated by the continuum model. The reason is that the additional decay $\Gamma^{(1)}$ associated with the sideband $\omega_a+\Omega$ is simply very slow compared to the dominant evolution determined by the high density of states near the band edge.  This explains why quasi-bound states near the lower band edge can be observed at finite velocities. However, since $\Gamma^{(1)}$ is non-zero, these bound states still decay on longer timescales.

To verify the validity of our analytic estimates we compare directly the predictions from the  effective model~\eqref{H_tot} with a numerical simulation of the full model~\eqref{H discrete}, which accounts for the exact time-dependence of the couplings. To do so we consider the specific example, where the local mode at each lattice site is represented by a Gaussian wavepacket
\begin{equation}\label{gaussian}
\bar u(z)=\frac{1}{\pi^{1/4}}\sqrt{\frac{a}{z_0}}e^{-\frac{z^2}{2z_0^2}},
\end{equation}
with a width $z_0\ll a$. This results in an averaged coupling
$\bar g\simeq g\sqrt{\frac{2z_0}{a}}\pi^{1/4}$.
For the simulation of the full model we assume $g/(2J)=0.1$ and $z_0/a=0.1$. This corresponds to $\bar g/(2J)\simeq 0.06$.
For these parameters we compare in Fig.~\ref{figsup1} d) and e) the excited state population predicted by the effective model and the full model for some of the values of $\delta$ and $v$ assumed in Fig.~\ref{fig:EmissionA}, Fig.~\ref{fig:EmissionB} b) and Fig.~\ref{fig:Transfer} a). For all these cases we see an excellent agreement between the exact and the effective evolution. More generally, we evaluate the evolution of the excited state population $p_e(t)$ for a single atom up to a final time $T_{\rm f}=50/J$ for a wide range of parameters $\delta$ and $v$. Then, the discrepancy parameter 
\begin{equation}
d= {\rm max} \{ |p_e^{\rm eff}(t)-p_e^{\rm full}(t)|, 0<t< T_f\},
\end{equation}
can be used to quantify the validity of the effective model.   In Fig.~\ref{fig:EmissionB} a) the black dashed line indicates the boundary set by the requirement $d\leq 0.1$. For most parameters, this boundary reproduces the condition identified in Eq.~\eqref{condition_main}. However, near $\delta\approx -2J$ the region of validity extends to very low velocities, which can be explained by the qualitative arguments given above.
We also see an additional kink at $v/\bar c \simeq 0.22$ and $\delta/(2J)=-1$. As shown in Fig.~\ref{figsup1} c), in this case also the $(\omega_a+\Omega)$-sideband hits a divergence in the photonic density of states and therefore induces a significant perturbation. 

In conclusion we find that, within the region of parameters delimited by the dashed line in  Fig.~\ref{fig:EmissionB} a),  the effective model~\eqref{H_tot} provides an accurate description of the dynamics and we can replace the modulated couplings $g(z_i(t))$ by the coupling $\bar g$ averaged over one period, $T=2\pi/\Omega$.

\section{Implementation}\label{Implementation}

While the above-described approach for realizing the slow-light Hamiltonian~\eqref{H_tot}  is very generic and can in principle be implemented with various photonic waveguide structures, the conditions for achieving a strong coupling between moving atoms and co-propagating photons are in practice very demanding.  In particular, for atomic velocities $v<10^4$ m/s and lattice constants of $a>10\,\mu$m, the condition $\bar c\sim v$ restricts the maximal bandwidth to $J/(2\pi)\approx  v/(4a\pi) < 80$ MHz, while it must still exceed the level of on-site disorder to avoid localization~\cite{Hernandez,Ciccarello_disorder,Douglas2015}. In addition, the atoms must be guided close to the surface of the waveguide such that the rate of emission into the waveguide, $\Gamma$,  exceeds the decay rate into free space, 
$\gamma_{a}$, as well as the photon loss rate, $\gamma_{p}$. 

In the following we first show in Sec.~\ref {sec:Fiber} how such extreme slow-light conditions can in principle be achieved using whispering gallery modes of a periodically modulated optical fiber. This setup is  motivated by the recent development of surface nanoscale axial photonic (SNAP) waveguides~\cite{SNAP}, where periodic photonic arrays with a high level of disorder control have already been demonstrated~\cite{Sumetsky2012}. In Sec.~\ref {micro} we then describe an alternative setup in the microwave regime, where many of the remaining experimental difficulties can be overcome by simply working with much larger wavelengths.

\subsection{Modulated optical waveguides}\label{sec:Fiber}
\begin{figure}
\includegraphics[width=0.48\textwidth]{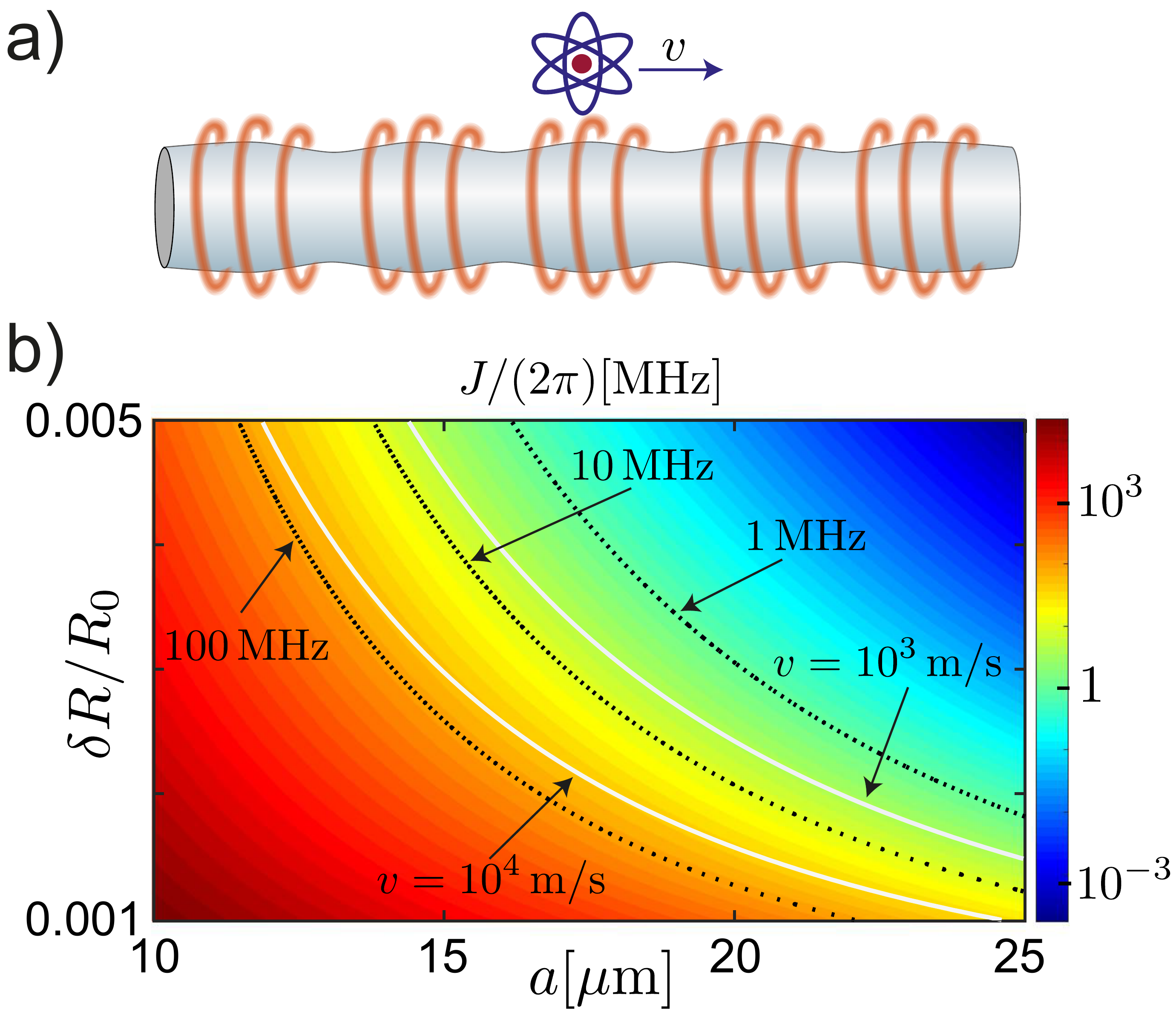}%
\caption{ a) Realization of an optical slow-light waveguide QED system using atoms coupled to the evanescent field of a silica fiber with a periodically modulated radius. Note that in this setup a natural reduction of the photonic speed already arises from the fact that the light does not propagate in a straight line, but rather around the circumference of the fiber. b) The tunneling amplitude $J/(2\pi)$ is plotted (in a logarithmic scale) as a function of the lattice constant $a$ and the radial variation $\delta R/R_0$. The white continuous lines indicate the condition $v=\bar c=2Ja$.  For this plot a cutoff frequency of  $\omega_e/(2\pi)\simeq 350$ THz ($\ell=180$) and a radius  $R_0=17.5\,\mu$m has been assumed. In order to avoid mixing with the neighboring transverse branches, the radial variation has been limited to $\delta R_0/R_0<1/\ell\simeq0.0056$.}\label{fig_optical_setup}
 \end{figure}

We consider a cylindric silica fiber with radius $R_0$ and refractive index $n\approx1.5$. As illustrated in Fig.~\ref{fig_optical_setup} a), the fiber supports optical modes, where light is guided around the circumference of the waveguide~\cite{SNAP,Lou} and couples evanescently to nearby atoms~\cite{ReitzPRL2013,YallaPRL2014}. For a given branch with a cutoff frequency 
 $\omega_e=\frac{\ell c}{nR_0}$ determined by the azimuthal quantum number $\ell$, the dispersion relation for small wavevektors $k$ along the z-direction is approximately given by
 \begin{equation}
 \omega_k \simeq  \omega_e +   \frac{\hbar k^2}{2m^*},
 \end{equation}
 where $m^*=\omega_e n^2\hbar/c^2\sim10^{-36}$ kg is the effective photon mass. Note that although this quadratic dispersion already gives rise to a vanishing group velocity near the band edge, it does not impose an upper bound on the propagation speed of the photons. Therefore, we assume in addition a small  periodic modulation of the fiber radius, 
  \begin{equation}\label{radial mod}
R(z)=R_0+\delta R\cos {\left(\frac{2\pi z}{a}\right)},
\end{equation} 
with a period $a$ that is large compared to the wavelength $\lambda_e=2\pi c/(n\omega_e)$. 
This modulation $R(z)$ results in a variation of the cutoff frequency, $\omega_e(z)=\omega_e R_0/R(z)$, which corresponds to an effective potential $V(z)=\hbar\omega_e (\delta R/R_0)\cos(2\pi z/a)$ for the photons propagating along $z$. As detailed in App.~\ref{AppCoptic}, the resulting dispersion relation $E_k=\hbar(\omega_k-\omega_e)$ and the Bloch waves $\psi_k(z)$ entering the model in Eq.~\eqref{H discrete} follow from the solutions of the effective Schr\"odinger equation~\cite{SNAP,Lou,Poll} 
\begin{equation}\label{effective Schroding}
\left[-\frac{\hbar^2}{2m^*}\frac{\partial^2}{\partial z^2}+ V(z)\right] \psi_k(z)= E_k \psi_k(z),
\end{equation}
and in the tight-binding limit $J$ can be obtained from the width of the first energy band. 
For the specific example of a fiber with radius $R_0=17.5\,\mu$m and $\omega_e/(2\pi)\simeq 350$ THz \cite{Poll,Poll2}
the resulting range of values for $J$ is plotted in Fig.~\ref{fig_optical_setup} b) for varying parameters $a$ and $\delta R$.

From Fig.~\ref{fig_optical_setup} b) we see that for lattice constants in the range of $a\sim10-20\,\mu$m we can obtain by this approach  a tunnel coupling of $J/(2\pi)\approx 80$ MHz together with an effective speed of light of $\bar c=10^4$ m/s. For the same fiber parameters we can estimate an atom-photon coupling of $g/(2\pi)\simeq 30-40$ MHz (see App.~\ref{AppCoptic}), in agreement with coupling constants measured for bottle resonators of similar dimensions~\cite{Poll2}. This translates into an averaged coupling of  $\bar g/(2\pi)\approx 10-20$ MHz, consistent with the requirements for the derivation of the effective continuum model.  At the same time, this coupling can exceed both the atomic decay rate, $\gamma_{a} /(2\pi)\simeq 2$ MHz, as well as photonic losses, $\gamma_{p}/(2\pi)\simeq 0.5$ MHz~\cite{Poll2}.

An ubiquitous problem of slow-light waveguides and coupled resonator arrays are unintended, fabrication-related variations of the local frequency, $\omega_e\rightarrow \omega_e(z)$. This additional random potential leads to localization of photons, when the bandwidth $J$ becomes too small. In App.~\ref{Disorder} we present a numerical study of our model in the presence of random on-site energies of magnitude $\epsilon$. From this study we estimate a tolerable level of disorder of $\epsilon/(2J)\leq 0.1$. This translates into maximal variations of the effective fiber radius of $\delta r \simeq R_0\epsilon/\omega_e\simeq 0.05 \dot A$. Note that this level of disorder control is achievable with in-situ tuning techniques, which have already been implemented for tunnel-coupled bottle resonator arrays~\cite{Sumetsky2012}, similar to the setup considered here. 

In summary these estimates show that, although challenging, a strong coupling of atoms and photons under the condition $\bar c\approx v$ can in principle be realized with modulated photonic waveguide structures. A remaining difficulty is the guiding of atoms at rather high velocities~\cite{Schrader,Schneeweiss} and at a distance of less than the optical wavelength above the waveguide. To overcome this problem it might be more favorable to consider waveguide QED systems in a much lower frequency regime.

\subsection{Slow-light waveguide QED with microwave photons and Rydberg atoms}\label{micro}

\begin{figure}
\includegraphics[width=0.48\textwidth]{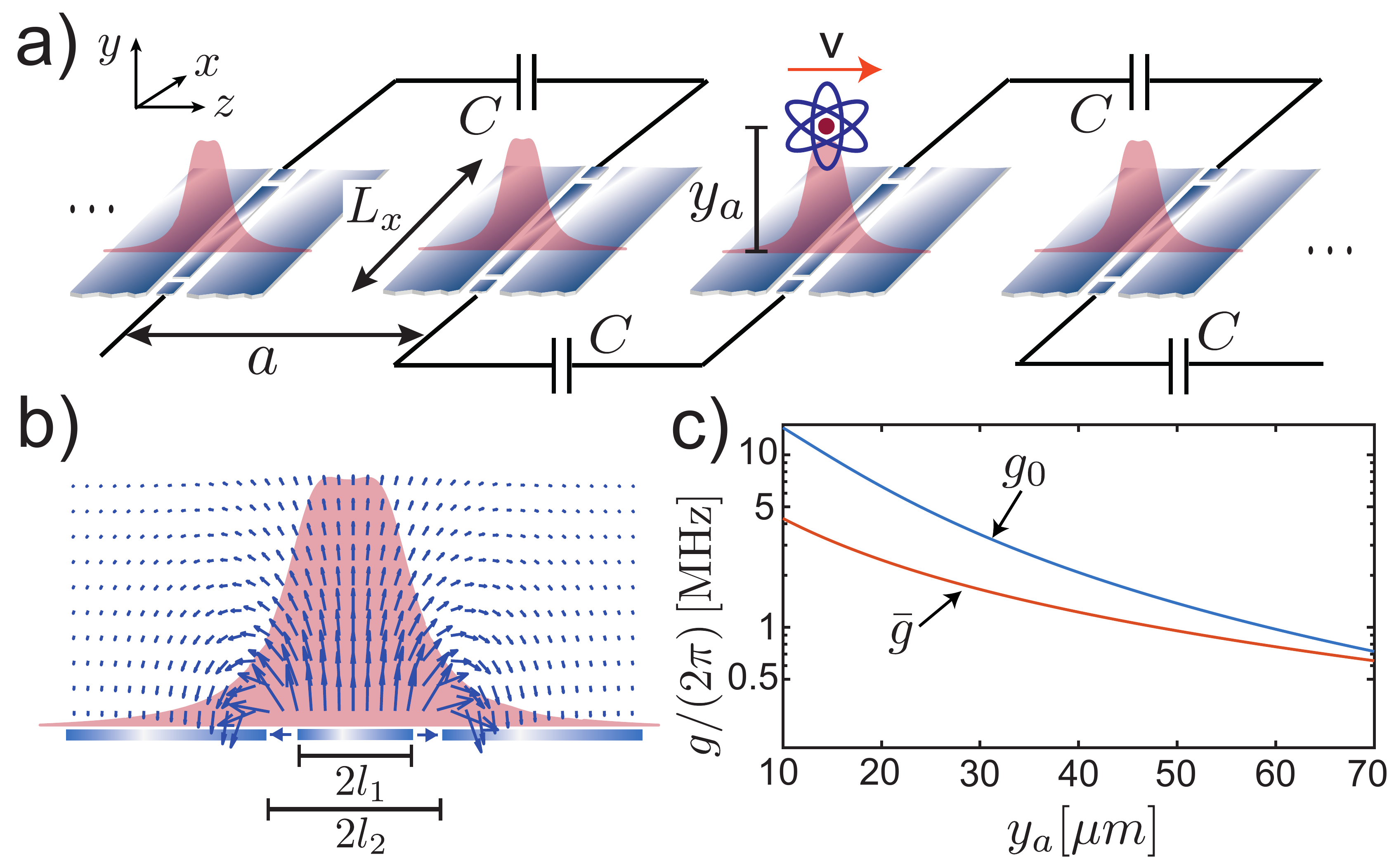}%
\caption{Slow-light waveguide QED with Rydberg atoms. a)  Rydberg atoms fly at a distance $y_a$ above an array of parallel CPW resonators coupled by capacitances $C$.  b) Cross section of a single resonator and sketch of the transverse field distribution. c) Atom-field coupling strength as a function of the atom-surface distance $y_a$ for the example described in the text. Here $g_0$ denotes the maximal coupling evaluated at $z=0$ and $\bar g$ the corresponding averaged  coupling, which is used in the effective continuum model~\eqref{H_tot}. See App.~\ref{AppC} for more details.}
\label{fig:Implementation}
\end{figure}

Let us now describe an alternative implementation shown in Fig.~\ref{fig:Implementation} a), where flying Rydberg atoms~\cite{petro,Hogan,HermannPRA2014,BeckAPL2016} are coupled to an array of coplanar waveguide (CPW) resonators.
For a length of $L_x\sim1$ cm the CPW resonators exhibit standing wave modes with frequencies $\omega_0=c\pi/L_x$ in the GHz regime, while in $z$-direction the electric field is strongly confined to a few tens of $ \mu$m determined by the size of the middle electrode, $2l_2$ [see Fig.~\ref{fig:Implementation} b)]. By considering an array of parallel resonators separated by $a\sim 100\,\mu$m, we obtain a closely spaced resonator array with a tunnel coupling $J$ that can be fully adjusted by additional capacitive couplings at the end-points~\cite{Houck,UnderwoodPRA2012}. 

As a specific example, we consider a microwave resonator with dimensions $L_x=0.5$ cm,  $l_1=10\, \mu$m, $l_2=15\,\mu$m and $\omega_0/(2\pi)\approx 30$ GHz. For $a= 200\,\mu$m and an atomic beam of velocity $v=10^4$ m/s~\cite{Hogan} we choose a capacitive coupling $J/(2\pi)=4$ MHz to obtain $v/\bar c\approx 1$. This bandwidth is still considerably above the level of on-site disorder achievable in large arrays of coupled microwave resonators~\cite{UnderwoodPRA2012}. By identifying two Rydberg states with neighboring principal quantum numbers $n$, i.e.  $|g\rangle=|n\rangle$  and $|e\rangle=|n+1\rangle$, we obtain a resonant coupling $\omega_a=\omega_{n+1}-\omega_n \simeq \omega_0$ for $n\simeq 60$. The resulting maximal atom-field coupling $g_0=d_{n,n+1}E_0(y_a)/\hbar$~\cite{petro,Hogan}, as well as the corresponding averaged coupling, $\bar g$, are plotted in Fig.~\ref{fig:Implementation} c) as a function of the atom surface distance, $y_a$. Here $d_{n,n+1}$ is the transition dipole moment and $E_0(y_a)$ the electric field per photon, which is evaluated in App.~\ref{AppC}. We see that already at convenient distances of $y_a\approx 50\,\mu$m we obtain  $\bar g\sim J/4$ and for the validity of the effective model we can choose any smaller value, e.g. $\bar g/J\simeq 0.1$, by adjusting the position of the beam along the $x$-direction. 

On the relevant timescale set by $\Gamma_c/(2\pi) \approx 400$ kHz, both the decay rate of the Rydberg states, $\gamma_{a}/(2\pi) \simeq 1$ kHz, and the photon loss rate $\gamma_{p}/(2\pi)\simeq 30$ kHz of a high-Q microwave resonator~\cite{Frunzio2005} are still negligible. For an array of $\sim500$ resonators with a total length of $L=10$ cm, the condition on the transverse spread during the flight time, $\Delta y/y_a<1$, requires a transverse cooling of the atom beam to a few mK, or alternatively, on-chip guiding systems for Rydberg atoms~\cite{Lancuba2016}. This shows that all the requirements for slow-light waveguide QED systems can be achieved in the microwave regime, using fabrication and atom-guiding techniques that  are currently developed.


\section{Conclusions}

In summary, we have described atom-photon interactions in the unconventional  regime, where the atomic velocities are comparable to the maximal speed of light in a waveguide structure. Under such conditions the atomic motion drastically changes photon emission and transfer processes and  induces directionalities and novel non-perturbative effects associated with the strong coupling of atoms and individual Cherenkov photons. We have shown how this combination of slow-light and strong coupling physics could potentially be reached with periodic waveguide structures in the optical and microwave regime. Specifically, we have introduced a new waveguide QED platform using Rydberg atoms flying above a coupled array of microwave resonators, where atom-light interactions under extreme slow-light conditions can be experimentally explored with present-day technology.

\section*{Acknowledgments}

The authors thank F. Ciccarello, D. Chang, A. Rauschenbeutel,  P. Schneeweiss, and J. Volz for valuable discussions. This work was supported by the Austrian Science Fund (FWF) through SFB FOQUS F40, DK CoQuS W 1210 and the START grant Y 591-N16.

\appendix

\section{Derivation and validity of the effective continuum model}\label{AppA}
In this appendix we provide additional details on the derivation of the effective continuum model in Eq.~\eqref{H_tot}. Starting from the full model given in Eq.~\eqref{H discrete}, we assume that at time  $t=0$ the $i$-th atom is located at the left border of one of the unit cells, i.e., $z_i(0)\equiv z_i=(n_1-1/2)a$, and the atom then traverses $l$ lattice sites following a classical trajectory $z_i(t)=z_i+v_i t$.  We here consider the case of very fast atoms where $\Omega_i=2\pi/T_i=2\pi |v_i|/a\gg g$.  In the interaction picture with respect to the bare atom and photon Hamiltonian, the integrated coupling between the atom and mode $a_k$ is then given by
\begin{equation}
\begin{split}
&G_k^l:= \int_0^{lT_i} dt'\,   g_k(z_i(t')) e^{ikz_i(t')}e^{i\delta_k t'}\\ &=\frac{g}{v_i} \int_{z_i}^{z_i+l a} dz\,   u_k(z) e^{ikz}e^{i\delta_k(z-z_i)/v_i},
\end{split}
\end{equation}
where $\delta_k=\omega_a-\omega_k$. We now use the fact that in the tight-binding limit,
\begin{equation}
u_k(z) \simeq \bar u(z) e^{-ikz}\qquad{\rm for} \,\,\,\, -a/2< z < a/2,
\end{equation}
where $\bar u(z)$ is approximately $k$-independent. Then, 
\begin{equation}\label{eq:GkSupp}
G_k^l =    \left[ \frac{g T_i}{a} \int_{-\frac{a}{2}}^{\frac{a}{2}} dz\,   \bar u(z)e^{i\delta_k z/v_i} \right]  \sum_{n=n_1}^{n_1+l-1}  e^{i k a n} e^{i\delta_k T_i (n+\frac{1}{2})} .
\end{equation}
To proceed  we neglect the variation $\sim e^{i\delta_k z/v_i}$ in the integral in Eq.~\eqref{eq:GkSupp}. This is valid for small detunings, $\delta_k\ll \Omega_i$, but also for strongly peaked $\bar u(z)$.  Under this approximation we can introduce the average coupling $\bar g$ given in Eq.~\eqref{gbar}
and obtain 
\begin{equation}
G_k^l \simeq  \bar g T_i  \times \sum_{n=n_1}^{n_1+l-1}  e^{i k a n} e^{i\delta_k T_i (n+1/2)} .
\end{equation}
We see that this result is just the discretized version of the integral 
\begin{equation}
G_k^l \simeq  \bar g \int_0^{l T_i}  dt' \,  e^{i k z_i(t')} e^{i\delta_k t'},
\end{equation}
which we would obtain for the same quantity evaluated with the effective model  given in Eq.~\eqref{H_tot}.

Let us now analyze  in more detail the range of validity for the continuum approximation. To do so we study  the evolution of the state $|\psi(t)\rangle= \big[\sum_ic_e^i(t) \sigma_+^i+ \sum_k  \, \psi(k,t) a^\dag_k\big]|g\rangle|{\rm vac}\rangle$. In the interaction picture with respect to $H_0=\sum_{i=1}^{N}  \hbar \omega_a |e\rangle_i\langle e| +\sum_{k}  \hbar \omega_k  \, a_k^{\dagger}a_k$ we obtain the following set of equations
\begin{equation}\label{sistime}
\begin{split}
&\frac{dc_e^i(t)}{dt}=-ig\sqrt{\frac{a}{L}}\sum_k \psi(k,t)u_{ki}(t)e^{i\tilde \delta_k^i t}e^{ikz_i}, \\
&\frac{d\psi(k,t)}{dt}=-ig\sqrt{\frac{a}{L}}\sum_j c_e^j(t)u^*_{kj}(t)e^{-i\tilde \delta_k^j t}e^{-ikz_j},
\end{split}
\end{equation}
where $\tilde\delta_k^i=\omega_a-(\omega_k-v_ik)$ and $u_{ki}(t)\equiv u_k(z_i(t))$.
After inserting the second equation into the first we obtain the following integro-differential equation
 \begin{equation}
 \begin{split}
\frac{dc_e^i(t)}{dt}=  -\frac{g^2a}{L} & \sum_j  \sum_k   e^{ik(z_i-z_j)}   u_{ki}(t) e^{i\tilde\delta_k^it}\\
 &\times   \int_0^t dt'   u^*_{kj}(t') e^{-i\tilde\delta_k^jt'} c_e^j(t')  .
 \end{split}
 \end{equation}
The $u_{ki}(t)$ are periodic in time and can be expanded  in a Fourier series $u_{ki}(t)=\sum_n u_{ki}^n e^{i \Omega_i n t}$. The resulting expression can be written as 
 \begin{equation}
 \begin{split}
 \frac{dc_e^i(t)}{dt} =&-\frac{g^2a}{L} \sum_{j}\sum_k   \sum_{n,m}  u^n_{ki}  [u^{m}_{kj}]^*   e^{ik(z_i-z_j)}    e^{i(\tilde\delta_k^i-\tilde \delta_k^j)t} \\
&\times e^{i(\Omega_i n-\Omega_j m) t} 
   \int_0^t dt'    e^{i\tilde\delta_k^j(t-t')} e^{i\Omega_j m (t-t')}  c_e^j(t') \\
 =& -\frac{g^2a}{L} \sum_j    \sum_k  u^0_{ki}  [u^{0}_{kj}]^*   e^{ik(z_i-z_j)}    e^{i(\tilde\delta_k^i-\tilde \delta_k^j)t} \\
 &\times \int_0^t dt'    e^{i\tilde\delta_k^j(t-t')} c_e^j(t') \, +\, {\rm rest}. 
 \end{split}
 \end{equation}
When evaluated under the same stroboscopic approximation as above, the first term in the second line of this equation simply corresponds to the evolution under the effective  model given in Eq.~\eqref{H_tot}.  A first set of corrections arises from terms with $n\neq 0$, but $m=0$. Such terms lead to additional oscillating contributions to $c_e^i(t)$. However, as long as $g\ll \Omega_i$, these corrections remain small and do not affect the long time behavior of $c_e^i(t)$. 

More crucial are terms with $n=m\neq 0$, which can generate non-oscillating contributions that affect the slowly varying dynamics of $c_e^i(t)$. To see this more explicitly, let us consider a single atom and a specific correction term
\begin{widetext}
 \begin{equation}
 \begin{split}
 \left.\frac{dc_e(t)}{dt}\right|_{n=m}&= -\frac{g^2a}{L}   \sum_k   |u^n_{k}|^2  \int_0^t dt'    e^{i\tilde\delta_k(t-t')} e^{i\Omega n (t-t')}  c_e(t')\approx  - \left[  \frac{\pi g^2a}{L}  \sum_k   |u^n_{k}|^2   \delta \left(\tilde\delta_k^i+\Omega_i n\right)  \right] c_e(t),
 \end{split}
 \end{equation}
 \end{widetext}
where in the second line we have made a Markov approximation and replaced the integral over $t$ by a $\delta$-function in frequency.  From this estimate we see that the higher order correction terms introduce additional decay channels. These occur when the  frequencies $\omega_a + n \Omega_i$ lie inside the tilted propagation band, as described in Sec.~\ref{Validity}. 

\section{Exact dynamics for an atom moving at the speed of light}\label{AppB}
 
In this appendix we derive the exact single excitation eigenstates for a stationary atom coupled to a photonic band with an exactly cubic dispersion relation, $\tilde\omega_k=\omega_c -J(ak)^3/3$, valid for $k\in(-\infty,\infty)$. This model is used as an approximation for the case $v=\bar c$, where such a dispersion relation is found in the co-moving frame. Note that compared to Eq.~\eqref{cubic} we here assume $k_c=0$ in order to simplify the notation. Since the bandwidth of this cubic model is infinite, all eigenstates are scattering states of the form $|\phi_k\rangle= \big[c_e^k \sigma_++ \int dk' \, \psi_k(k') a^\dag_{k'}\big]|g\rangle|{\rm vac}\rangle$ with energy $E_k=\hbar\tilde\omega_k$. From the Schr\"odinger equation $\tilde H|\phi_k\rangle=E_k|\phi_k\rangle$ we obtain  \begin{equation}\label{sistbu1}
\begin{split}
&c_e^k(\tilde \omega_k-\omega_a)=\bar g\sqrt{\frac{a}{2\pi}}\int dk'\psi_k(k')\,, 
\end{split}
\end{equation}
while the
Lippmann-Schwinger equation for the photonic wavefunction reads~\cite{Palma,Eco}
\begin{equation}\label{psik}
\psi_k(k')=\sqrt{\frac{a}{2\pi}}\frac{\bar g c_e^k}{\tilde\omega_k- \tilde\omega_{k'}+ i\epsilon}+\delta(k-k')\mbox{.}
\end{equation}
Here the limit $\epsilon \rightarrow 0^+$ is assumed to obtain the correct boundary conditions. By  inserting Eq.~\eqref{psik} back into Eq.~\eqref{sistbu1} the atomic amplitude can be written in a closed form as
\begin{equation}
c_e^k=\sqrt{\frac{a}{2\pi}}\frac{\bar g}{ \tilde\omega_k-\omega_a-\bar g^2I_0(\tilde\omega_k)}.
\end{equation}
Here we have introduced the integral  
\begin{equation}\label{Integral}
\begin{split}
I_{z}(\tilde\omega_k)&=\frac{a}{2\pi}\int \frac{e^{ik'z} dk'}{\tilde\omega_k-\tilde\omega_{k'}+ i\epsilon}=-\frac{\sgn(k)\sqrt{3}+i}{2Ja^2k^2}\\ &\times \left(e^
{-i\frac{kz}{2}-\frac{\sqrt{3}}{2}|kz|}+\theta(-z)e^{i\frac{2}{3}\pi \sgn[k]+ikz}\right),
\end{split}
\end{equation}
where $\sgn(k)$ is the sign function and $\theta(z)$ is the Heaviside function with $\theta(z=0)=0$. 
 The scattering states can then be written in momentum space as
\begin{equation}
|\phi_k\rangle=\left[c_e^k\sigma_+ + a_k^\dag+\sqrt{\frac{a}{2\pi}}\int dk'\frac{\bar g c_e^k a_{k'}^\dag}{\tilde\omega_k-\tilde\omega_{k'}+ i\epsilon}\right]|g\rangle|{\rm vac}\rangle\mbox{,}
\end{equation} 

To evaluate the photonic wavefunction $\psi_k(z)=\frac{1}{\sqrt{2\pi}}\int dk'  \psi_k(k') e^{ik'z}$ in position space, 
we make again  use of the integral in Eq.~\eqref{Integral} and we obtain
 \begin{widetext}
\begin{equation}
\psi_k(z)=\frac{1}{\sqrt{2\pi}}\left[e^
{ikz}+\gamma_{k}\left(e^
{-i\frac{kz}{2}-\frac{\sqrt{3}}{2}|kz|}+\theta(-z)e^{i\frac{2}{3}\pi \sgn[k]+ikz}\right)\right],
\end{equation}
\end{widetext}
with a scattering amplitude 
\begin{equation}
\gamma_{k}=\frac{\bar g^2I_0(\tilde\omega_k)}{\tilde\omega_k-\omega_a-\bar g^2I_0(\tilde\omega_k)}.
\end{equation}
%
%
Note that these scattering states are very a-typical since independent of the direction of the incoming wavevector $k$, the scattered wave vanishes at $z\rightarrow +\infty$, while it is finite at $z\rightarrow -\infty$. This can be understood from the particularity of the assumed dispersion relation, which always leads to negative group velocities.


Using the exact scattering solutions we can investigate the spontaneous emission of an initially excited atom. In this case the atom-field dressed state $|\psi(t)\rangle= \big[c_e(t) \sigma_++ \int dk  \, \psi(k,t) a^\dag_k\big]|g\rangle|{\rm vac}\rangle$ can be expressed in terms of the eigenstates of the system according to
\begin{equation}
|\psi(t)\rangle=\int dk \, (c_e^{k})^*e^{-i\tilde\omega_k t}|\phi_k\rangle,
\end{equation}
and we obtain  $c_e(t)=\langle e|\psi(t)\rangle$ or the emitted photonic wave packet
\begin{equation}
\psi(z,t)=\langle z|\psi(t)\rangle=\int dk\, \psi_k(z) (c_e^{k})^*e^{-i\tilde\omega_k t},
\end{equation}
in terms of integrals over $k$. To derive a simpler approximate result for the atomic population it is more convenient to consider instead the equations of motion for $c_e(t)$ and $\psi(k,t)$. In a frame rotating with $\omega_c=\omega_a-\delta$ they are given by
\begin{eqnarray}
\dot c_e^k &=& - i \delta c_e(t)  - i \bar g\sqrt{\frac{a}{2\pi}}\int dk'\psi_k(k',t),\\
\dot \psi(k,t) &=&  i \frac{J}{3}(ka)^3  \psi(k,t)  - i \bar g \sqrt{\frac{a}{2\pi}} c_e(t).
\end{eqnarray}
By performing a Laplace transformation we obtain for the atomic amplitude
\begin{equation}\label{laplace}
c_e(s)=\frac{1}{s+i\delta+\bar g^2I(s)},
\end{equation}
where
 \begin{equation}
 I(s)=i\frac{a}{2\pi}\int \frac{dk}{J(ka)^3/3+is}=(9Js^2)^{-1/3}.
 \end{equation}
 %
%
For vanishing detuning, i.e., $\omega_a=\omega_c$, we can further  rewrite $c_e(s)$ as
 \begin{equation}
 \begin{split}
 c_e(s)=&\frac{1}{s+i\bar g\left(\frac{s}{9J}\right)^{1/6}}+\frac{1}{s-i\bar g\left(\frac{s}{9J}\right)^{1/6}}\simeq \\
&\frac{1}{s+i\Omega_c+\Gamma_c/2}+\frac{1}{s-i\Omega_c+\Gamma_c/2},
\end{split}
\end{equation}
where in the last step we made a single pole approximation. Under this approximation the inverse Laplace transform of $c_e(s)$ results in a damped cosine function as given in Eq.~\eqref{pop_approx}  with the  the characteristic oscillation frequency $\Omega_c$ and the damping rate $\Gamma_c$ given in Eq.~\eqref{coefficients}.

 \section{Whispering gallery modes in a modulated optical fiber}\label{AppCoptic}
We consider a cylindrical fiber as described in Sec.~\ref{sec:Fiber} with a small radial modulation as specified in Eq.~\eqref{radial mod}. The eigenmodes $\vec \Phi_k(\vec r)$ of the electric field in the waveguide with frequency $\omega_k$ and wavevector $k$ along the $z$ direction are solutions of the Helmholtz equation
\begin{equation}
\left[\Delta+ n^2(\vec r) \frac{\omega_k^2}{c^2}\right]\vec\Phi_k(\vec r)=0,
\end{equation}
where $n(\vec r)$ is the refractive index that assumes the values $n(\vec r)=n\simeq 1.47$ inside the fiber and $n(\vec r )=1$ outside.
In view of the cylindrical symmetry we change to polar coordinates and make the ansatz $\vec \Phi_k(\vec r)=\vec\chi_\ell (r,z) \psi_k(z)e^{i\ell \phi}$, where $\ell$ is the azimuthal mode index. The radial component $\vec\chi_\ell (r,z)$ satisfies the radial Helmholtz equation for a given $R(z)$ and corresponding eigenfrequency $\omega_e(z)=\frac{\ell c}{\bar n c_r R(z)}$. Here $\bar n$ denotes the effective refractive index averaged over the radial mode profile and $c_r$ is a correction factor that accounts for the finite radial width of the mode function~\cite{Poll}. Note that these small corrections have been omitted in the main text for simplicity. For a TM mode, i.e., $\vec\chi_\ell = \chi_{\ell,z}  \vec e_z$, the unnormalized radial mode functions are given by~\cite{Lou,Poll,Poll2,Yar}
\begin{equation}
  \begin{split}
&\chi_{\ell,z} (r,z) =J_\ell\left(k_0(z)n r\right)\mbox{,     } r\le R(z)\\
&\chi_{\ell,z} (r,z)=\left(\frac{J_\ell(k_0(z)n R(z))}{Y_\ell(k_0(z) R(z))}\right)Y_\ell\left(k_0(z) r\right)\mbox{,     } r > R(z)
\end{split}
\end{equation} 
where $J_\ell(x)$ and $Y_\ell(x)$ are  Bessel functions of the first and second kind
and  $k_0(z)=\omega_e(z) n/c$. 
Given a certain frequency $\omega_e$ the fiber radius and the azimuthal number $\ell$ are related  by the resonance condition
\begin{equation}
n\frac{J_\ell'(k_0(z)n R(z))}{J_\ell(k_0(z)n R(z))}-\frac{Y_\ell'(k_0(z) R(z))}{Y_\ell(k_0(z) R(z))}=0.
\end{equation}

By making a paraxial approximation, valid for $\delta R/a\ll 1$, we neglect the derivatives of $\vec\chi_\ell (r,z)$ with respect to $z$ and we obtain  
\begin{equation}
\left[\frac{\partial^2}{\partial z^2} + \frac{\bar n^2}{c^2} (\omega_k^2-\omega^2_e(z))\right] \psi_k(z) =0. 
\end{equation}
After some small manipulations and introducing the eigenenergies $E_k=\hbar(\omega_k-\omega_e)$ this equation corresponds to the effective Schr\"odinger equation given in Eq.~\eqref{effective Schroding}. From the numerical solutions of this equation we estimate the tunnel coupling $J$ from the width of the first energy band, i.e.,  
$
J=({\rm max}\{\omega_k\}-{\rm min}\{\omega_k\})/4$.

In the absence of the periodic modulation the atom field coupling strength evaluated at the surface of the fiber reads 
\begin{equation}
g=d\sqrt{\frac{\omega_k}{2\hbar \epsilon_0aA}} |\vec\chi_\ell (R_0)|.
\end{equation}
Here $\epsilon_0$ is the vacuum permittivity, $d\simeq2\times10^{-29}$ Cm is the dipole moment of a $^{133}$Cs atom and $A$ is the mode area defined as~\cite{Lou,Poll}
\begin{equation}
\begin{split}
A=&\pi\left(n^2 \int_{0}^{R_0}dr rJ_\ell^2(k_0nr)+\right. \\
&\left.\frac{J_\ell^2(k_0nR_0)}{Y_\ell^2(k_0R_0)}\int_{R_0}^{\infty}dr rY_\ell^2(k_0r)\right).
\end{split}
\end{equation}
  The  radial modulation gives an approximate  quadratic potential for the photons  in each lattice site. Thus, the local modes  can be represented by Gaussian wavepackets of the form given in Eq.~\eqref{gaussian}. This leads to an effective coupling $\bar g\simeq g \pi^{1/4}\sqrt{2z_0/a}$ as assumed for all our estimates.

 \section{Disorder}\label{Disorder}

To evaluate the influence of disorder on the present slow-light waveguide system, we consider a random frequency offset $\delta \omega_l$ at each lattice site $l$. In momentum space this corresponds to the Hamiltonian
 \begin{equation}
H_{\rm dis}=\hbar \sum_{k,k'}f_{k,k'}a_k^{\dagger}a_{k'},
\end{equation}
where $f_{k,k'}=\frac{a}{L}\sum_l e^{i(k-k')l}\delta \omega_l$, which we add in our numerical simulation to the original Hamiltonian given in Eq.~\eqref{H_tot}. For numerical simulations the $\delta \omega_l$ are chosen randomly from a uniform distribution within the interval $[-\epsilon/2,\epsilon/2]$.

\begin{figure}
\includegraphics[width=0.48\textwidth]{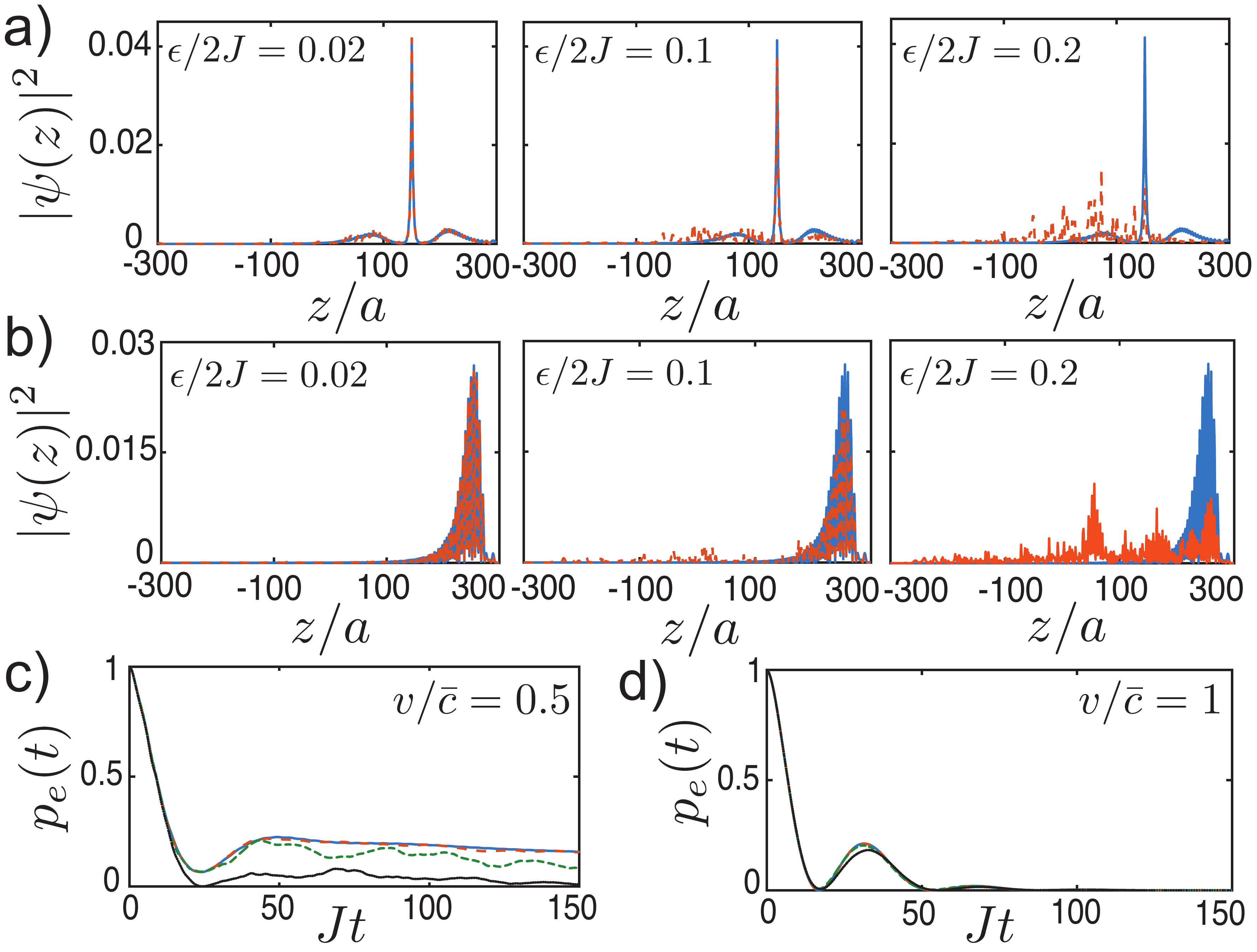}%
\caption{The effect of disorder on the system dynamics. The red dashed lines in the plots in a) and b) show the emitted photon wavepacket (evaluated at $t=110/J$) for increasing levels of disorder and for the parameters a) $v/ \bar c=0.5$, $\delta/(2J)=-0.5$ and b) $v/\bar c=1$,  $\delta/(2J)=-\pi /2$. For comparison the case without disorder is represented by the blue solid line.  The plots in c) and d) show the decay of the excited state population for  $\epsilon/(2J)=0$ (continuous line), $\epsilon/(2J)=0.02$ (long dashes), $\epsilon/(2J)=0.1$ (short dashes) and $\epsilon/(2J)=0.2$  (dotted line). In all plots a coupling of  $\bar g/(2J)=0.1$ is assumed.  }
\label{fig:disorder}
\end{figure}

In Fig.~\ref{fig:disorder} a) and c) we plot the atomic population and the wave function of the emitted photon for the case $v/ \bar c=0.5$ and $\delta/(2J)=-0.5$ and for different values of the disorder strength $\epsilon$. From the plots we see that the system dynamics is almost unaffected by the disorder up to values of about $\epsilon/(2J)=0.1$. Above this value the signatures of a co-moving bound state are washed out. In Fig.~\ref{fig:disorder} b) and d) we plot the analog results for   the critical coupling conditions $v/\bar c=1$ and $\delta=-\pi J$. Here we find that the atomic decay is even less affected by disorder. However, for $\epsilon/(2J)>0.1$ the emitted wavefunction  becomes again significantly distorted, which would affect the dynamics of excitation transfer processes. 

In conclusion, we find that all the effects described in the main text are robust with respect to disorder up to a strength of the order of $\epsilon/(2J)\simeq 0.1$.  For further discussions on the effect of disorder in coupled resonator waveguides see also Ref.~\cite{Ciccarello_disorder} and the supplementary material of \cite{Douglas2015}. 

\section{Estimate of the atom-field coupling strength for a CPW resonator}\label{AppC}
For a CPW resonator as shown in Fig.~\ref{fig:Implementation} b) the electric field of the fundamental mode is given by 
\begin{equation}
\vec{E}(x,y,z)=\left[E_z(z,y)\vec{e}_z+E_y(z,y)\vec{e}_y\right]\cos{\left(\frac{\pi x}{L_x}\right)}.
\end{equation}

To estimate the transverse electric field profile we consider the limit of an infinitely flat CPW resonator with an additional ground plate located at a distance $h$ below the central track. Further, if the width of the two outer  ground planes is larger than twice the gap between them, they can be treated as infinitely extended~\cite{Pon}.  With this assumption, an exact solution for the electric field above the central electrode of the CPW resonator can be obtained using conformal mapping techniques~\cite{Gilli}. In particular,  the two transverse field components are given by $E_z(z,y)={\rm Im}\{E(t)\}$ and $E_y(z,y)={\rm Re}\{E(t)\}$, where $t=z+iy$ and for $y>0$ 
\begin{equation}
E(t)= \frac{E_0}{\sqrt{\left(\frac{t^2}{l_1^2}-1\right)\left(\frac{t^2}{l_2^2}-1\right)}}.
\end{equation}
Finally, for the evaluation of the coupling strength $g$ we need the electric field per photon, $E_0=\sqrt{\frac{\hbar \omega_0}{2\epsilon_0 V_r}}$, where $V_r$ the mode volume of the resonator which can be found by imposing the normalization condition
\begin{equation}
\epsilon_0\frac{L_x}{2}\left[\int_{-\infty}^{\infty}d z\int_{-h}^{\infty}d y\left(E_z^2(z,y)+E_y^2(z,y)\right)\right]=\frac{\hbar\omega_0}{2}.
\end{equation}
For the parameters considered in the main text and $h<5 \mu$m,  the volume below the central electrode is negligible  compare to the mode volume above and only the latter must be considered for the evaluation of the field strength $E_0$.


\end{document}